\documentclass[12pt]{article}
\def\nabstar#1{\nabla\kern-0.5pt\smash{\raise 4.5pt\hbox{$\ast$}}
               \kern-4.5pt_{#1}}

\def\drvstar#1{\partial\kern-0.5pt\smash{\raise 4.5pt\hbox{$\ast$}}
               \kern-5.0pt_{#1}}

\def\newline{\relax\ifhmode\null\hfil\break\else\nonhmodeerr@\newline\fi}
\def\frac#1#2{{#1\over#2}}
\def\text#1{{\hbox{\rm #1}}}

\newcommand{\beq}{\begin{equation}}
\newcommand{\eeq}{\end{equation}}
\newcommand{\bea}{\begin{eqnarray}}
\newcommand{\eea}{\end{eqnarray}}
\def\Id{ \mbox{1\hspace{-1.2mm}I} }
\def\EQ{\hspace{-2mm} &=& \hspace{-2mm}}

\def\BA{\begin{eqnarray}}
\def\EA{\end{eqnarray}}
\def\BAN{\begin{eqnarray*}}
\def\EAN{\end{eqnarray*}}

\def\nn{\nonumber\\}

\def\tr{\mbox{tr}}

\def\gm5{\gamma_5}

\input epsf.sty
\newdimen\psfigsize
\def\psfigure#1 #2 #3 #4 #5{
    \begin{figure}[tbh]
      \begin{center}
      \vbox{
        \null\vskip-0.2in\hskip#2
        \epsfxsize=#1
        \epsfbox{#4}
        \vskip -0.3in
        \caption {#5 \label{#3}}
        \vskip 0.0 true in plus 0.3 true in
      }
      \end{center}
   \end{figure}
}
\usepackage{graphicx}
\begin{document}
\thispagestyle{empty}
\begin{flushright}
NTUTH-04-505C \\
June 2004 \\
\end{flushright}
\bigskip\bigskip\bigskip
\vskip 2.5truecm
\begin{center}
{\LARGE {$ N (N^*) $ and $ \Delta (\Delta^*) $ on the lattice
         }}
\end{center}
\vskip 1.0truecm
\centerline{Ting-Wai Chiu and Tung-Han Hsieh}
\vskip5mm
\centerline{Department of Physics and} 
\centerline{National Center for Theoretical Sciences at Taipei} 
\centerline{National Taiwan University, Taipei 106, Taiwan}
\vskip 1cm
\bigskip \nopagebreak \begin{abstract}
\noindent

We investigate the mass spectrum of Nucleon and Delta (and its
counterparts with strange and charm), and their 
excited states, in quenched lattice QCD with exact chiral symmetry.
For each light baryon, we use 23 masses to determine 
the coefficients of the mass formula in quenched chiral 
perturbation theory. By chiral extrapolation to  
$ m_\pi = 135 $ MeV, we obtain  
$ M_N = 958(26) $ MeV, $ M_{N^*} = 1553(42) $ MeV,
$ M_{\Delta} = 1216(32) $ MeV and
$ M_{\Delta^*} = 1611(17) $ MeV, which are identified with 
$ N(939) P_{11} $, $ N(1535) S_{11} $,  
$ \Delta(1232)P_{33} $ and $ \Delta(1620)S_{31} $ respectively.
Further, we directly measure the masses of $ \Omega^{-} $,  
$ M_{\Omega} = 1648(60) $ MeV, and its  
excited state, $ M_{\Omega^*} = 1935(48) $ MeV;
as well as the triply charmed baryon $ \Omega_{ccc}^{++} $,   
$ M_{\Omega_{ccc}^{++}} = 4931(22) $ MeV, 
and its excited state, $ M_{{\Omega^{++}_{ccc}}^*} = 5185(35) $ MeV.

\vskip 1cm
\noindent PACS numbers: 11.15.Ha, 11.30.Rd, 12.38.Gc

\noindent Keywords: Nucleon, Delta, Excited baryons, Charmed baryons, 
Lattice QCD

\end{abstract}
\vskip 1.5cm 
\newpage\setcounter{page}1

One of the objectives of lattice QCD is to compute the hadron masses
nonperturbatively from the first principles. 
For hadrons composed of charm and strange quarks 
(i.e., without $ u, d $ light quarks), their masses can be 
directly measured on presently accessible lattices.
However, for hadrons containing $ u, d $ light quarks,  
the performance of the present generation of computers is still
quite remote from what is required for computing their masses
at the physical scale (e.g., $ m_\pi \simeq 135 $ MeV), on a lattice
with enough sites in each direction such that the discretization
errors as well as the finite volume effects are
both negligible comparing to the statistical ones.
Nevertheless, even with lattices of moderate sizes, lattice QCD
can determine the values of the parameters in the hadron mass
formulas of the (quenched) chiral perturbation theory.
Then one can use these formulas to evaluate the hadron masses at the
physical scale, as well as to determine the quark masses.

In this paper, we study spin 1/2 and spin 3/2 baryons containing 
three quarks of the same mass\footnote{In this paper, we work in the 
isospin limit, $ m_u = m_d $.}, and obtain their masses for 30 quark
masses in the range $ 0.03 \le m_q a \le 0.8 $ (i.e., from   
$ m_s/2 $ to $ m_c $). 

For spin 3/2 baryons composed of three strange quarks ($ \Omega^{-} $), 
and three charm quarks ($ \Omega_{ccc}^{++} $), as well as 
their excited states, we can measure their masses directly on the lattice.
Here the strange quark bare mass ($ m_s a = 0.06 $) and charm quark bare 
mass ($ m_c a = 0.8 $) are fixed by requiring the masses of 
the corresponding vector mesons to agree with the masses of 
$ \phi(1020) $ and $ J/\psi (3097) $. 

For baryons composed of light quarks, we use 23 masses to 
determine the coefficients in the baryon mass formula in 
quenched chiral perturbation theory \cite{Labrenz:1996jy}
\bea
\label{eq:baryon_qChPT}
M = c_0 + c_1 m_\pi 
    + c_L m_\pi^2 \ln( m_\pi ) 
    + c_2 m_\pi^2  
    + c_3 m_\pi^3 + O(m_\pi^4 \ln m_\pi)
\eea  
Then, extrapolating to the physical pion mass $ m_\pi = 135 $ MeV, 
we obtain the masses of  
$ N(1/2^+), N^*(1/2^-), \Delta(3/2^+) $, and $ \Delta^*(1/2^-) $ respectively. 
For our data, we find that we have to set $ c_1 = 0 $, otherwise 
the errors of some coefficients would become unacceptable ($>100\%$). 
For $ \Delta^* $, we also have to set $ c_3 = 0 $.
Evidently, if one can measure more baryon masses in 
the chiral regime ($ m_\pi \ll 2 \sqrt{2} \pi f_\pi $), then the 
coefficients can be determined more precisely.
However, one should limit $ m_\pi L > 3 \sim 4 $, otherwise 
the baryon masses would suffer from finite size effects, 
which in turn would lower the reliability of the resulting coefficients. 
For our lattice of size $ 20^3 \times 40 $ at $ \beta = 6.1 $, 
we find that $ m_q a \ge 0.03 $ gives $ m_\pi L > 4 $ in the 
spatial direction, which should be sufficient to suppress the 
finite size effects.

The interpolating operators for Nucleon and $ \Delta $ can be 
written as 
\bea
\label{eq:nucleon}
N_{x\alpha} = 
\epsilon_{abc} 
 {\bf u}^T_{x\alpha b} (C\gamma_5)_{\alpha\beta} {\bf d}_{x\beta c}
{\bf u}_{x\alpha a}    
\eea 
and 
\bea
\label{eq:delta}
\Delta_{x\alpha;\mu} =     
\epsilon_{abc} 
{\bf d}^T_{x\alpha b} (C\gamma_5 \gamma_\mu )_{\alpha\beta} {\bf d}_{x\beta c}
{\bf d}_{x\alpha a} \ ,    
\hspace{2mm} \mu=1,2,3
\eea 
where ${\bf u}$ and ${\bf d}$ denote the quark fields,
the superscript $ T $ denotes the transpose of the Dirac spinor,
$ C $ is the charge conjugation operator,
$ \epsilon_{abc} $ is the completely antisymmetric tensor,
and $ x $, $ \{ a, b, c \} $ and $ \{ \alpha, \beta \} $
denote the lattice site, color, and Dirac indices respectively.
For $ \Delta (\Delta^*) $, we average over 
$ \mu = 1, 2, 3 $ to increase the statistics.
Note that the ``diquark" operator in (\ref{eq:delta}) transforms like 
a vector, thus the even parity state of (\ref{eq:delta}) can 
overlap with $ \Delta(3/2^+) $, in which the quarks are in S-wave; 
while its odd parity state can overlap with 
$ \Delta^{*} (1/2^-) $, in which the quarks are in P-wave.

Now it is straightforward to work out the baryon propagator
in terms of quark propagators. 
In lattice QCD with exact chiral symmetry,
quark propagator with bare mass $ m_q $ is of the form 
$ (D_c + m_q )^{-1} $ \cite{Chiu:1998eu}, 
where $ D_c $ is exactly chirally symmetric
at finite lattice spacing. 
In the continuum limit, $ (D_c + m_q)^{-1} $ reproduces
$ [ \gamma_\mu ( \partial_\mu + i A_\mu ) + m_q ]^{-1} $. 
For optimal domain-wall fermion \cite{Chiu:2002ir} 
with $ N_s + 2 $ sites in the fifth dimension,   
\BAN
D_c \EQ 2m_0 \frac{1 + \gamma_5 S(H_w)}{1 - \gamma_5 S(H_w)}, \\
S(H_w) \EQ \frac{1 - \prod_{s=1}^{N_s} T_s}
                {1 + \prod_{s=1}^{N_s} T_s}, \\
 T_s \EQ \frac{1 - \omega_s H_w }{1 + \omega_s H_w},  \hspace{4mm}
 H_w = \gamma_5 D_w,  
\EAN
where $ D_w $ is the standard Wilson Dirac operator plus a negative 
parameter $ -m_0 $ ($0<m_0<2$), and $ \{ \omega_s \} $ are a set of 
weights specified by an exact formula such that $ D_c $ 
possesses the optimal chiral symmetry \cite{Chiu:2002ir}. Since 
\BAN
( D_c + m_q )^{-1} 
= (1-rm_q)^{-1} [ D^{-1}(m_q) - r ],  \hspace{4mm} r = \frac{1}{2m_0}
\EAN
where   
\BAN
D(m_q) = m_q + (m_0 - m_q/2)[ 1 + \gamma_5 S(H_w) ], 
\EAN
thus the quark propagator can be obtained by solving 
the system $ D(m_q) Y = \Id $ with nested conjugate 
gradient \cite{Neuberger:1998my},
which turns out to be highly efficient (in terms of the precision 
of chirality versus CPU time and memory storage) if the inner 
conjugate gradient loop is iterated with Neuberger's double pass 
algorithm \cite{Neuberger:1998jk}.
For more details of our scheme of computing quark propagators, 
see Ref. \cite{Chiu:2003iw}.

We generate 100 gauge configurations with Wilson gauge action
at $ \beta = 6.1 $ on the $ 20^3 \times 40 $ lattice.
Fixing $ m_0 = 1.3 $, we project out 16 low-lying eigenmodes of 
$ |H_w| $ and perform the nested conjugate gradient in the complement
of the vector space spanned by these eigenmodes. For  
$ N_s = 128 $, 
the weights $ \{ \omega_s \} $ are fixed with $ \lambda_{min} = 0.18 $ 
and $ \lambda_{max} = 6.3 $, 
where $ \lambda_{min} \le \lambda(|H_w|) \le \lambda_{max} $
for all gauge configurations.    
For each configuration, quark propagators are computed
for 30 bare quark masses in the range $ 0.03 \le m_q a \le 0.8 $, 
with stopping criteria $ 10^{-11} $ and $ 2 \times 10^{-12} $
for the outer and inner conjugate gradient loops respectively.
Then the norm of the residual vector of each column of the quark propagator 
is less than $ 2 \times 10^{-11} $  
\BAN
|| (D_c + m_q ) Y - \Id || < 2 \times 10^{-11},  
\EAN
and the chiral symmetry breaking due to finite $ N_s $ is 
less than $ 10^{-14} $,
\BAN
\sigma = \left| \frac{Y^{\dagger} S^2 Y}{Y^{\dagger} Y} - 1 \right|
< 10^{-14},
\EAN
for every iteration of the nested conjugate gradient. 

After the quark propagators have been computed, we first 
measure the pion propagator and its time correlation function, and 
extract the pion mass ($ m_\pi a $) and the pion decay constant  
($ f_\pi a $). 
With the experimental input $ f_\pi = 132 $ MeV,  
we determine the inverse lattice spacing $ a^{-1} = 2.21(3) $ GeV. 
(Our procedure has been described in Ref. \cite{Chiu:2003iw}.)

Next we compute the nucleon propagator 
$ \langle N_{x\alpha} \bar N_{y\delta} \rangle $ 
and its time correlation function $ C(t) $.
Then the average of $ C(t) $ over gauge configurations is 
fitted by the usual formula 
\BAN
\label{eq:ctBL}
\frac{1+\gamma_4}{2} ( Z_{+} e^{-m_{+} a t}
                              -Z_{-} e^{-m_{-}a (T-t)}) 
   +\frac{1-\gamma_4}{2} ( Z_{+} e^{-m_{+}a (T-t)}
                              -Z_{-} e^{-m_{-}a t})
\EAN   
where $ m_\pm $ are the masses of even and odd parity states. 
Thus, one can use parity projector $ ( 1 \pm \gamma_4 )/2 $
to project out two amplitudes,
\BAN 
A_{+}(t) &\equiv& Z_{+} e^{-m_{+}a t} -Z_{-} e^{-m_{-}a (T-t)}, \nn 
A_{-}(t) &\equiv& Z_{+} e^{-m_{+}a (T-t)} -Z_{-} e^{-m_{-}a t}. 
\EAN

Now the problem is how to extract $ m_{\pm} $ from $ A_{\pm} $ respectively. 
Obviously, for sufficiently large $ T $, there exists 
a range of $ t $ such that, in $ A_{\pm} $, the contributions due to 
the opposite parity state are negligible. 
Then $ m_{\pm} $ can be extracted by a single exponential fit to $ A_{\pm} $,  
for the range of $ t $ in which the effective mass 
$ m_{eff}(t) = \ln(A_{\pm}(t)/A_{\pm}(t+1)) $ attains a plateau.    
On the other hand, if $ T $ is not so large, then it may turn out 
that the heavier mass, say $ m_- $ (assuming $ m_- > m_+ $), could not be 
easily extracted from $ A_- $ due to the non-negligible contributions of 
the (lowest lying) even parity state. Further, if $ T $ is too small, 
then one even has difficulties to extract the mass of the lowest lying state. 
For our lattice with $ T = 40 $, it is sufficiently large to extract both 
$ m_{+} $ and $ m_{-} $ from $ A_{+} $ and $ A_{-} $ respectively. 
 
\begin{figure}[htb]
\begin{center}
\hspace{0.0cm}\includegraphics*[height=12cm,width=10cm]{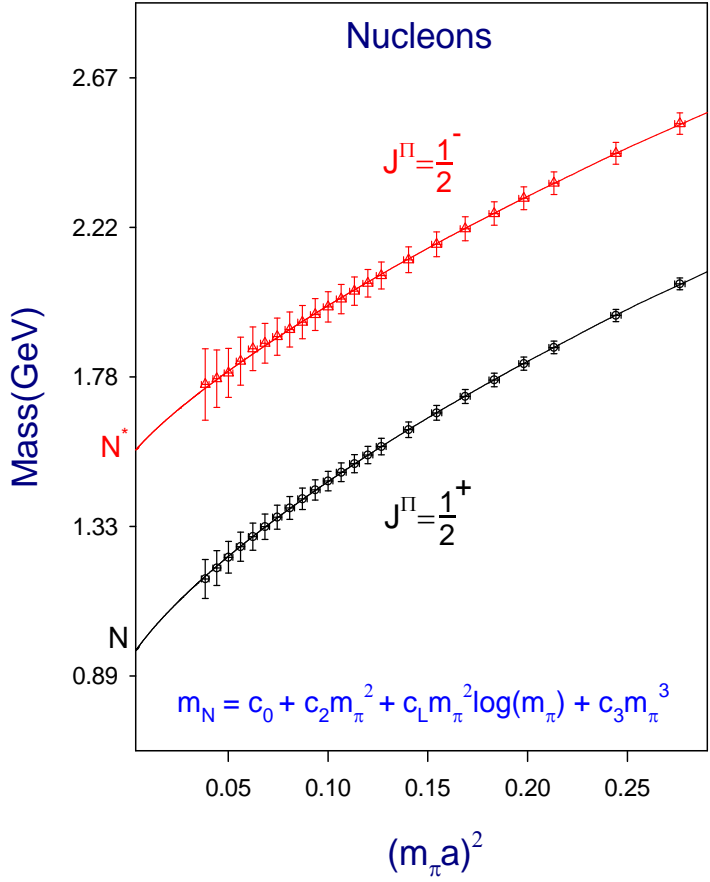}
\caption{The nucleon masses versus the pion mass square, 
for $ N(J^\Pi = 1/2^+) $ 
and $ N^*(J^\Pi = 1/2^-) $ respectively. 
The solid lines are fits to (\ref{eq:baryon_qChPT}) with $ c_1 = 0 $. 
}
\label{fig:nucleon_mass}
\end{center}
\end{figure}

In Fig. \ref{fig:nucleon_mass}, the nucleon masses of even and odd parity 
states are plotted versus $ (m_\pi a)^2 $, for $ m_\pi $ smaller  
than the chiral cutoff $ \Lambda_\chi =  2 \sqrt{2} \pi f_\pi $. 
For the $ J^\Pi = 1/2^+ $ ground state nucleon, 
the 23 nucleon masses can be fitted by 
\bea
\label{eq:N_fit}
M_N a \EQ 0.417(13) 
- 1.350(168) \times (m_\pi a)^2 \ln( m_\pi a) 
+ 0.727(134) \times (m_\pi a)^2 \nn
&& + 0.458(144) \times (m_\pi a )^3 
\eea
with $ \chi^2$/d.o.f. $ < 0.003 $. 
At the physical pion mass $ m_\pi = 135 $ MeV (the $y$-axis), (\ref{eq:N_fit})
gives $ M_N = 958(26) $ MeV, 
which is naturally identified with $ N(939) P_{11} $. 

For the $ J^\Pi = 1/2^- $ excited nucleon, the 23 masses can be fitted by
\bea
\label{eq:N*_fit}
M_{N^*} a \EQ 0.689(21) - 0.949(291) \times (m_\pi a)^2 \ln( m_\pi a) 
+ 0.904(236) \times (m_\pi a)^2 \nn
&& + 0.211(250) \times (m_\pi a )^3 
\eea
with $ \chi^2$/d.o.f. $ < 0.007 $. 
At $ m_\pi = 135 $ MeV, (\ref{eq:N*_fit}) gives
$ M_{N^*} = 1553(42) $ MeV, which is identified with 
$ N(1535) S_{11} $.


\begin{figure}[htb]
\begin{center}
\hspace{0.0cm}\includegraphics*[height=12cm,width=10cm]{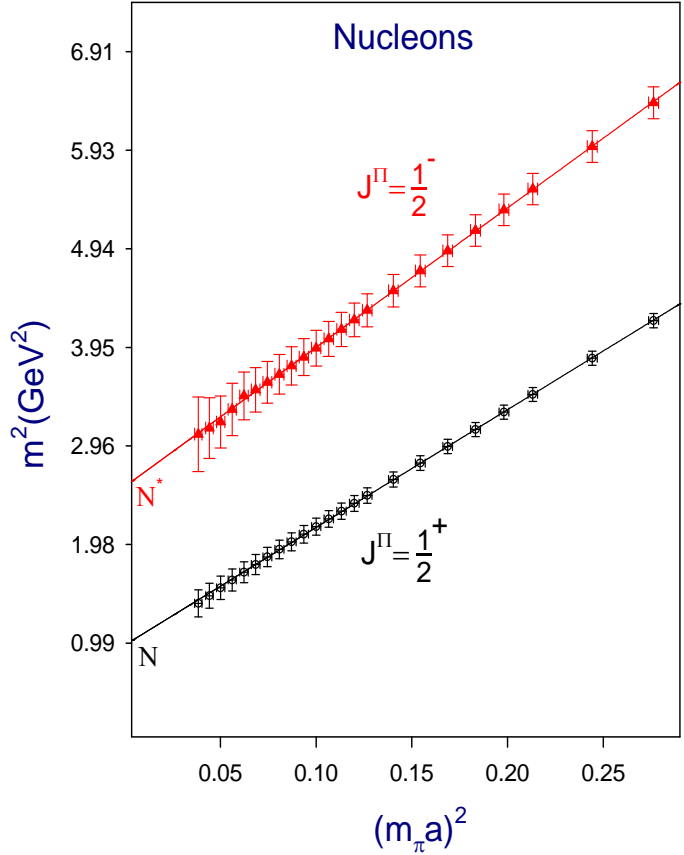}
\caption{The nucleon mass square versus the pion mass square, 
for $ N(J^\Pi = 1/2^+) $ 
and $ N^*(J^\Pi = 1/2^-) $ respectively.  
}
\label{fig:mN2_mpi2}
\end{center}
\end{figure}

At this point, it is instructive to examine whether our data of nucleon mass 
can accomodate the linear term $ c_1 m_\pi $.  
To this end, we take the square of the nucleon mass,     
then the higher order terms are further suppressed, and the linear term 
(if any) would become more prominent, especially at small $ m_\pi a $. 
In Fig. \ref{fig:mN2_mpi2}, we plot $ (M_N a)^2 $ and $ (M_{N^*} a )^2 $ 
versus $ (m_\pi a )^2 $, which turn out to be well fitted by
\bea
\label{eq:mN2_fit}
(M_{N} a)^2 \EQ 0.196(9) + 2.388(55) \times (m_\pi a)^2 \\
\label{eq:mN*2_fit}
(M_{N^*} a )^2 \EQ 0.5174(192) + 2.829(124) \times (m_\pi a)^2
\eea
Obviously, the term $ c_1 \sqrt{(m_\pi a)^2} $ does not exist 
in our data of nucleon masses, otherwise, we would see deviations 
from the fitted lines at small $ (m_\pi a) $.

Note that (\ref{eq:mN2_fit}) and (\ref{eq:mN*2_fit}) 
give $ M_N = 1001 (22) $ MeV and $ M_N^{*} = 1607 (30) $ MeV 
at $ m_\pi = 135 $ MeV, in good agreement with the masses of 
$ N(939) P_{11} $ and $ N(1535) S_{11} $. 
This seems to justify the ansatz, namely,       
if the number of baryon mass data points is too small 
(e.g., less than 10) and/or the baryon masses are measured 
only at large quark masses, then they may not feasible for 
fitting the mass formula (\ref{eq:baryon_qChPT}), however,   
they still could be used for  
the linear fit: $ (M a)^2 = A + B (m_\pi a)^2 $, which may turn out to   
provide a good extrapolation to the physical $ m_\pi $. 

For recent quenched lattice QCD studies of $ N $ and $ N^* $, see Refs. 
\cite{Lee:1998cx}-\cite{Brommel:2003jm}.

\begin{figure}[htb]
\begin{center}
\hspace{0.0cm}\includegraphics*[height=12cm,width=10cm]{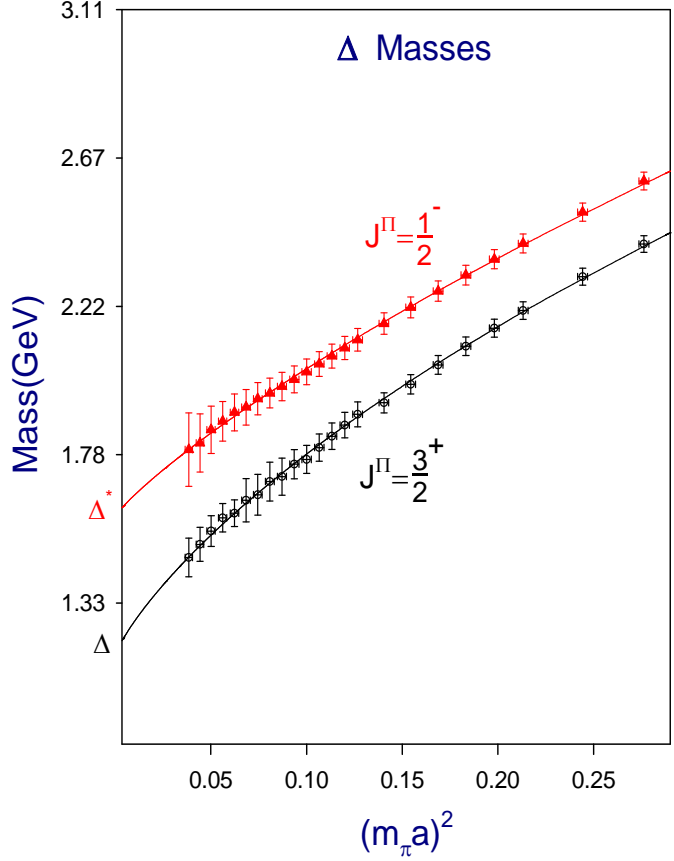}
\caption{The $ \Delta $ masses versus the pion mass square,
for $ \Delta (J^\Pi = 3/2^+) $
and $ \Delta^*(J^\Pi = 1/2^-) $ respectively.
The solid lines are fits to (\ref{eq:baryon_qChPT}), with $ c_1 = 0 $
for $ \Delta $, and with $ c_1 = c_3 = 0 $ for $ \Delta^* $.
}
\label{fig:delta}
\end{center}
\end{figure}

\begin{figure}[htb]
\begin{center}
\hspace{0.0cm}\includegraphics*[height=12cm,width=10cm]{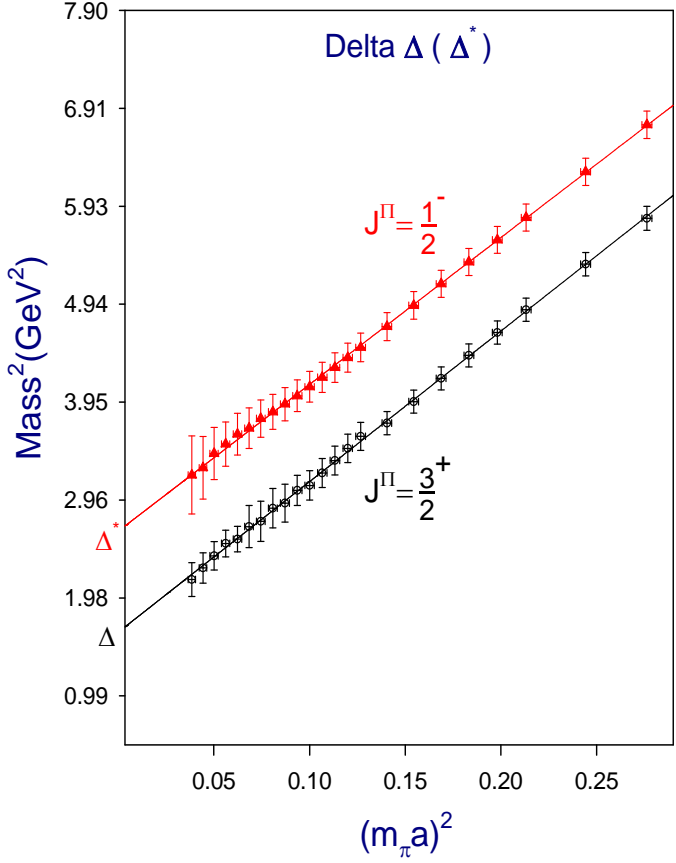}
\caption{The Delta mass square versus
the pion mass square,
for $ \Delta (J^\Pi = 3/2^+) $
and $ \Delta^*(J^\Pi = 1/2^-) $ respectively.
The solid lines are linear fits.
}
\label{fig:mD2_mpi2}
\end{center}
\end{figure}

Next we compute the time correlation function of $ \Delta $, 
and extract the masses of even and odd parity states. 

In Fig. \ref{fig:delta}, we plot the masses of $ \Delta $   
versus $ m_\pi^2 $ for even and odd parity states respectively.
The lowest lying state has even parity, which implies that it has  
$ J^\Pi = 3/2^+ $ since its three quarks are in S-wave, 
and symmetric in spin. 
Using (\ref{eq:baryon_qChPT}) with $ c_1 = 0 $ to fit the 23 masses of 
$ \Delta $, we obtain    
\eject
\bea
\label{eq:delta_fit}
M_{\Delta} a \EQ 0.530(16) 
- 1.685(218) \times (m_\pi a)^2 \ln( m_\pi a) 
+ 0.552(176) \times (m_\pi a)^2 \nn
&& + 0.682(187) \times (m_\pi a )^3 
\hspace{10mm} (J^{\Pi} = 3/2^+)  
\eea
At $ m_\pi = 135 $ MeV, (\ref{eq:delta_fit})
gives $ M_{\Delta} = 1216(32) $ MeV, 
which is identified with $ \Delta(1232)P_{33} $. 

For the negative parity state $ \Delta^* $, 
its quarks are in P-wave, thus its $ J^\Pi = 1/2^- $. 
Using (\ref{eq:baryon_qChPT}) with $ c_1 = c_3 = 0 $ to fit
the 23 masses of $ \Delta^* $, we obtain    
\bea
\label{eq:delta*_fit}
M_{\Delta^*} a \EQ 0.717(8) 
- 0.724(34) \times (m_\pi a)^2 \ln( m_\pi a) 
+ 1.156(12) \times (m_\pi a)^2 
\eea
At $ m_\pi = 135 $ MeV, (\ref{eq:delta*_fit})
gives $ M_{\Delta^*} = 1611(17) $ MeV, which is identified with  
$ \Delta(1620)S_{31} $. This is the first lattice QCD determination 
of the mass of $ \Delta(1620)S_{31} $. 

Again, we check the ansatz of baryon mass square versus pion mass square. 
In Fig. \ref{fig:mD2_mpi2}, we plot $ (M_\Delta a)^2 $ and 
$ (M_{\Delta^*} a )^2 $ versus $ (m_\pi a )^2 $. 
They can be fitted by the straight lines:
\bea
\label{eq:mD2_fit}
(M_{\Delta} a)^2 \EQ 0.330(13) + 3.078(87) \times (m_\pi a)^2 
\hspace{10mm} [J^{\Pi} = 3/2^+]  \\
\label{eq:mD*2_fit}
(M_{\Delta^*} a )^2 \EQ 0.536(17) + 3.000(109) \times (m_\pi a)^2
\hspace{10mm} [J^{\Pi} = 1/2^-]
\eea
At $ m_\pi = 135 $ MeV, they give $ M_\Delta = 1292(25) $ MeV and 
$ M_{\Delta^*} = 1635(26) $ MeV, in good agreement  
$ \Delta(1232)P_{33} $ and $ \Delta(1620)S_{31} $ respectively. 

Finally, we turn to the heavy baryons which can be measured directly
on our lattice. Now replacing the down quark in $ \Delta^- $ with 
the strange quark, we have the $ \Omega^{-} $. Similarly,  
replacing the up quark in $ \Delta^{++} $ with the charm quark, 
we obtain the triply charmed baryon $ \Omega_{ccc}^{++}   
(J^\Pi = 3/2^+)$,  
which so far has not been discovered in high energy experiments. 

To determine the bare masses of strange quark and charm quark, 
we extract the mass of vector meson from the time correlation function  
\BAN
C_\rho (t) = \frac{1}{3} \sum_{\mu=1}^3 \sum_{\vec{x}} 
\tr\{ \gamma_\mu (D_c + m_q)^{-1}(0,x) \gamma_\mu
     (D_c + m_q)^{-1}(x,0) \}
\EAN
At $ m_q a = 0.06 $, the vector meson has mass $ M a = 0.4638(32) $, 
which gives $ M = 1025(7) $ MeV, in good agreement with 
the mass of $ \phi(1020) $. Thus, at $ m_q a = 0.06 $, 
$ \Delta^{-} $ becomes $ \Omega^{-} $, and its mass can be extracted 
from the time correlation function. Our result is 
$ M_\Omega a = 0.746(27) $, which yields   
\bea
\label{omega}
M_{\Omega^-} = 1648(60) \mbox{MeV} 
\eea
in good agreement with the mass of $ \Omega^{-}(1672) $. 

Similarly, at $ m_q a = 0.80 $, the vector meson has mass   
$ M a = 1.3830(15) $, which gives $ M = 3056(3) $ MeV, in 
good agreement with the mass of $ J/\Psi(3097) $. 
Thus, at $ m_q a = 0.80 $, $ \Delta^{++} $ actually is $ \Omega_{ccc}^{++} $, 
the triply charmed baryon with $ J^\Pi = 3/2^+$, which 
has not been observed in experiments. However, we can determine 
its mass from the time correlation function, and our result is 
$ M a = 2.23(1) $, which gives   
\bea
\label{eq:delta_c}
M_{\Omega_{ccc}^{++}} = 4931(22) \mbox{MeV}  
\hspace{10mm} [\mbox{quark content=(ccc)}, \ J^\Pi = 3/2^+] 
\eea
Also, for the corresponding $ J^\Pi = 1/2^- $ excited baryons, 
our results are 
\bea
\label{eq:omega*}
M_{\Omega^*} \EQ 1935(48) \mbox{MeV}  
\hspace{10mm} [\mbox{quark content=(sss)}, \ J^\Pi = 1/2^-] \\
\label{eq:delta_c*}
M_{{\Omega_{ccc}^{++}}^*} \EQ 5185(35) \mbox{MeV}  
\hspace{10mm} [\mbox{quark content=(ccc)}, \ J^\Pi = 1/2^-] 
\eea
Note that (\ref{eq:delta_c})-(\ref{eq:delta_c*}) are the first
lattice QCD predictions for the masses of these baryons.  
Even though we have not estimated the errors due to quenched approximation, 
discretization, and finite lattice size, we suspect that the resulting error  
in each mass of (\ref{eq:delta_c})-(\ref{eq:delta_c*}) is 
less than 80 MeV. A possible candidate for the 
$ \Omega^{*} $ in (\ref{eq:omega*}) 
could be the $ \Omega^{-}(2250) $ \cite{Hagiwara:fs}.

Finally we transcribe the bare masses of strange quark ($ m_s a = 0.06 $) and 
charm quark ($ m_c a = 0.80 $) to $ \overline{\mbox{MS}} $ at $ \mu = 2 $ GeV. 
Using the lattice renormalization constant 
$ Z_m = Z_s^{-1} $ 
(where $ Z_s $ is the renormalization constant for $ \bar\psi \psi $)
to one loop order \cite{Alexandrou:2000kj}, we obtain
\bea
\label{eq:mass_s}
m_s^{\overline{\mbox{MS}}}(\mbox{2 GeV}) \EQ \mbox{114(9) MeV} \\
\label{eq:mass_c}
m_c^{\overline{\mbox{MS}}}(\mbox{2 GeV}) \EQ \mbox{1520(38) MeV} 
\eea
where the errors are estimates based on the errors of the  
vector meson masses with respect to $ \phi(1020) $ and $ J/\Psi(3097) $.




This work was supported in part by the National Science Council,
ROC, under the grant number NSC92-2112-M002-023.
T.W.C. would like to thank David Lin for discussions.

\bigskip
\bigskip

\vfill\eject

\end{document}